\documentclass[lettersize,journal]{IEEEtran}
\IEEEoverridecommandlockouts
\usepackage{cite}
\usepackage{amsmath,amsthm,amssymb,amsfonts}
\usepackage{algorithmic}
\usepackage{graphicx}
\usepackage{epstopdf}
\usepackage{textcomp}
\usepackage{xcolor}
\usepackage{balance}
\usepackage{tikz}

\newcommand{\BE}{\begin{equation}}
	\newcommand{\EE}{\end{equation}}

\def\BibTeX{{\rm B\kern-.05em{\sc i\kern-.025em b}\kern-.08em
		T\kern-.1667em\lower.7ex\hbox{E}\kern-.125emX}}
\begin{document}
	\title{Integrating Mediumband with Emerging Technologies: Unified Vision for 6G and Beyond Physical Layer} 
\author{Dushyantha A Basnayaka,~\IEEEmembership{Senior Member,~IEEE}, and Abdulla Firag
\thanks{Dr. Basnayaka is an Assistant Professor currently based in Dublin, Ireland. Prior to that he was with The University of Edinburgh, UK. Dr. Firag is currently a consultant with Asian Development Bank. Prior to that he was with Dhiraagu|the first and the largest telecommunication service provider in Maldives. Corresponding Author: \textit{d.basnayaka@ieee.org}.}
%
}
\markboth{This work has been submitted to IEEE for publication.}%
{Shell \MakeLowercase{\textit{et al.}}: A Sample Article Using IEEEtran.cls for IEEE Journals}
\maketitle
\begin{abstract}
		In this paper, we present a vision for the physical layer of 6G and beyond, where emerging   physical   layer   technologies   integrate   to   drive   wireless   links   toward mediumband operation, addressing a major challenge: deep fading, a prevalent, and perhaps the most consequential, obstacle in wireless   communication   link   performance.   By   leveraging   recent   insights   into wireless   channel   fundamentals   and   advancements   in   computing, multi-modal sensing, and AI, we articulate how reflecting surfaces (RS), sensing, digital twins (DTs), ray-tracing, and AI can work synergistically to lift the burden of deep fading in future wireless communication networks. This refreshingly new approach promises transformative improvements in reliability, spectral   efficiency,   energy   efficiency,   and   network   resilience,   positioning   6G   for truly superior performance.
\end{abstract}
	\vspace{0mm}
	\section{Introduction}
	\indent On the broad shoulders of some individuals, teams, and in some cases of organizations, the wireless communication has risen from an obscure luxury in the 90's to an everyday commodity in just over 30 years. Wireless communication that is fast and reliable has brought people closer and driven tremendous economic growth globally. For the continual economic growth by increasing and facilitating economic activities among people, the enabling technologies like wireless communication should be made to be as ubiquitous, reliable, fast, and also efficient as possible.\\ 
	\indent The widely known 5G is the state-of-the-art in modern digital wireless communication for mass communication, which was started to be rolled out in early 2020. The world is currently poised, driven by the rapidly growing application areas and trends, to introduce the next major advancement to the global wireless communication: 6G. Many experts have been speculating about, and some are exploring, the potential enabling technologies for 6G physical layer (PHY) since early 2019.\\
	\indent The use of gigantic MIMO, which is a scaled-up version of massive MIMO \cite{CuiDai22}, further densification of cells, use of new frequencies in upper mid-band (i.e., 7.125 – 24.25 GHz) and sub-THz and THz bands (broadly from 100 GHz to 10 THz) \cite{Kang24, Madanayake24}, integrating sensing and communication, the introduction of, as they call it, ``reflecting surfaces'' or ``intelligent metasurfaces'' to physically manipulate the propagation environment \cite{WuZhang20, Mayur24}, and the exploitation of AI/ML techniques \cite{Qin19, Alkhateeb20} and digital twins \cite{XLin23, Alkhateeb23} are the notable emerging technologies for 6G over the years. To this mixture, we now also have an emerging concept called ``\textit{mediumband wireless communication}'', which is rather a new operating point or condition for wireless communication systems \cite{Bas2023}. It has been shown that wireless systems operating in the mediumband can achieve high data rates, afford more interference, and perhaps more importantly suppress deep fading giving rise to significantly superior and robust wireless communication systems \cite{BasMag2024}.\\
	\indent The 6G enabling technologies are powerful in their own ways, but as we are transitioning from exploration to systematization, it is not yet clear how these technologies could co-exist smoothly and effectively to form a unified, coherent, and backward compatible physical layer for 6G and beyond. There is an urgent need for new ideas to introduce/revise use cases, where the individual capabilities of these emerging technologies add constructively to build a superior physical layer for 6G and beyond in terms of reliability, spectral and energy efficiency, robustness and resiliency.
\begin{figure*}[t]
	\centerline{\includegraphics[scale=0.9]{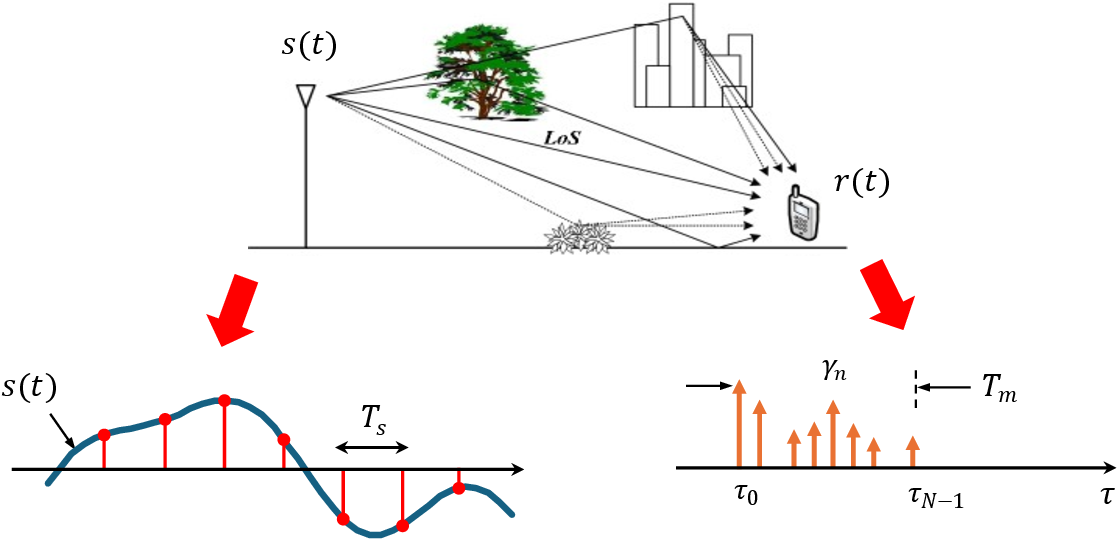}}
	\caption {On the top is a typical propagation environment, where $s(t)$ is the transmit signal and $r(t)$ is the receive signal. On the bottom-left is a depiction of a typical $s(t)$. In digital wireless communication, not all parts of $s(t)$, but only the points regularly separated in time (i.e. RED dots), carry information. This separation time is the symbol period, $T_s$. On the bottom-right is a depiction of a multipath delay profile as an impulse train, where the $x$--coordinate and the height of the impulses represent the excess delay, $\tau_n$, and the strength, $\gamma_n$, of the corresponding MPC respectively, where $T_m$ is a suitable measure of the time spread of MPCs like maximum excess delay\cite{Sarkar03}. A wireless communication system, whose $T_m$ and $T_s$ approximately satisfy the condition: $T_m < T_s < \lambda T_m$, which is represented by the GREEN region in Fig. \ref{fig:fig0}, is said to be operating in the mediumband.}\label{fig:fig21}
	\vspace{0mm}
\end{figure*}
\vspace{-2mm}
\subsection{Lessons From 5G}
Massive MIMO, cell densification, and the introduction of new frequency bands are at the heart of the 5G PHY, which is an OFDM-based broadband system. Being OFDM-based means that the wireless signals between transmitters (TXs) and receivers (RXs) in 5G are broadband waveforms generated using the multi-carrier technology: OFDM. These are brilliant and powerful technologies individually, but many have varied views on their combined effect in practice.\\
\indent 5G encountered significant adoption challenges. Unlike previous generations, among the concerns raised by service providers is that 5G was launched without a compelling business case or high-demand use case, as 4G’s speed and latency already satisfied most existing applications at the time. Consequently, many service providers introduced 5G primarily as a marketing strategy to position themselves as technological leaders rather than to meet specific user needs. Even after its launch, high deployment costs and a lack of commercially viable use cases hindered 5G’s rollout, with many providers struggling to justify the substantial infrastructure investments required. Additionally, 5G necessitates costly, dedicated hardware incompatible with much of the 4G infrastructure, particularly at the access layer, and primarily uses higher-frequency bands to support faster speeds. Although this higher-spectrum bandwidth enables greater data rates, it reduces coverage, demanding more base stations and further increasing costs.\\
\indent These lessons emphasize the need for well-defined, compelling use cases, enabling technologies, and an architecture that address 5G’s deployment and cost challenges when planning for 6G and beyond physical layer. Furthermore, 6G should be able to address the practicalities like enabling service providers to benefit from their investments within a reasonable time frame. Moreover, disagreement among stakeholders and regions on use cases and enabling technologies risk standard and market fragmentation endangering the success of 6G.    
\vspace{-3mm}
\subsection{Beyond 5G}
Multi-carrier modulation introduced in 4G exploits both time and frequency dimensions effectively for high rate and highly reliable physical layer. In addition to time and frequency dimensions, using massive MIMO, 5G elevated the use of the spatial dimension to new heights. While plans are underway to expand the use of time, frequency and spatial dimensions further in 6G, the consensus is that harnessing the potential of wireless propagation environment for the benefit of 6G and beyond is the immediate next frontier in physical layer wireless communication. How could 6G enabling technologies smoothly co-exist while harnessing the potential of wireless propagation environment, and what is the sweet spot?\\         
\indent This paper, exploiting a theoretical framework introduced in \cite{Bas2023,BasMag2024}, presents a unified vision to harness the potential of wireless environment, where several powerful technologies like reflecting surfaces (RS), multi-modal sensing, digital twins, ray-tracing computations, and machine/deep learning, and their synergies, join hands to form a physical layer for 6G and beyond.  This architecture envisions a scenario where every constituent link\footnote{Note that this link could be in a sub-6GHz massive MIMO system or in a upper mid-band cell free MIMO network or in a sub-THz wireless personal area network (WPAN) or in a non-terrestrial-network (NTN) or in a cognitive radio system or in a relay communication system or in a narrowband low-power wide area network (LPWAN), or in a wireless ad-hoc network.} in the network receive help from aforementioned technologies, not only for physical layer functions like beam management, blockage prediction, link adaptation, and CSI acquisition, but also more importantly to operate in the mediumband. By operating in the mediumband, wireless links, and in turn the entire network, can suppress the most problematic effect caused by the propagation environment: deep fading. Such operation is expected to create a favourable knock-on impact on innumerable aspects in physical and other layers.\\
\indent We start with an elaboration as to what these new terms: ``mediumband'' and ``mediumband wireless communication'' really mean, and how to make sure a wireless communication system operates in the mediumband in practice. Subsequently, our vision of integrating mediumband with other emerging technologies is explained in detail with some concluding remarks in the following sections. 
\section{What is Mediumband?}
\begin{figure}[t]
	\centerline{\includegraphics*[scale=0.7]{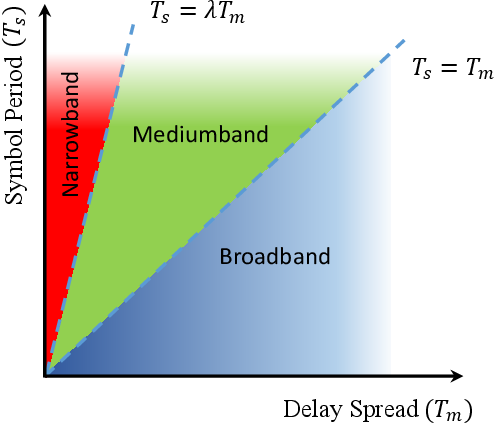}}
	\caption{\cite{Bas2023,BasMag2024} Main regions on the $T_mT_s$--plane representing three classes of wireless channels, where the exact value of the constant, $\lambda$ is dependent on the exact definition of $T_m$. For instance, if $T_m$ is the maximum excess delay, $\lambda \approx 10$, but if $T_m$ is defined differently $\lambda$ would be different. Importantly, whatever the lines of separation of different regions, mediumband on the $T_mT_s$--plane refers to the region between narrowband and broadband regions where something called ``\textit{the effect of deep fading avoidance}'' is most prominent. Henceforth, $\lambda=10$ is assumed.}\label{fig:fig0}
	\vspace{-2mm}
\end{figure}

	The term ``\textit{mediumband}'' does NOT refer to any particular band of frequencies in the electromagnetic spectrum like microwave band, mmwave band or terahertz band, but to a class of wireless channels with unique properties, which may hold the key to solve some of the biggest challenges in modern wireless communication, if used cleverly \cite{BasMag2024}. For those, who are not sufficiently acquainted with the concept of wireless channels, mediumband can simply be regarded as a ``\textit{set of constraints}'' defined in terms of two important parameters in wireless communication: the first is the delay spread and the other is the symbol period. \textbf{The readers must distinguish mediumband from another closely sounding term: mid-band, which just refers to a band of frequencies in the EM spectrum roughly from 3.5GHz to 25GHz, and conceptually mid-band has nothing to do with mediumband.} What do these symbol period and delay spread really mean in the context of digital wireless communication?\\
	\indent Considering a typical communication link with a single transmitter (TX) and a single receiver (RX), the receive signal $r(t)$ at the RX is a random mixture of delayed and attenuated versions of the transmit signal $s(t)$. This $r(t)$ can be mathematically given in baseband equivalent form by:
	\begin{align} \label{eq11}
		r(t) &= \underbrace{\sum _{n=0}^{N-1} \gamma_{n} s\left ({t-\tau _{n}}\right )}_{r'(t)} + w(t),
	\end{align}
	where $\gamma _{n}$ and $\tau _{n}$ respectively are the complex path gain and path delay of the $n$th multipath component (MPC). Here, $w(t)$ is additive noise and $N$ is the number of MPCs. The actual values of $\gamma_{n}$ and $\tau _{n}$ are dependent on the propagation environment between the TX and the RX. As shown in Fig. \ref{fig:fig21}, in $s(t)$, only the signal points regularly separated in time carry digital data, and this separation time is the symbol period ($T_s$). The delay spread ($T_m$) is a measure of the time spread of the MPCs. It is typically a spatially averaged measure, and is dependent on the propagation environment, whereas $T_s$ is dependent on the wireless hardware. Every wireless communication system occupies a unique point defined in terms of $T_m$ and $T_s$ on the $T_mT_s$--plane in Fig. \ref{fig:fig0}, where the mediumband constraint is shown in GREEN, which mathematically is: $T_m < T_s < 10T_m$. What does mediumband wireless communication or communicating in the mediumband really mean?     
\vspace{0mm}
\section{Communicating In the Mediumband}	
\begin{figure*}[t]
	\centerline{\includegraphics*[scale=0.86]{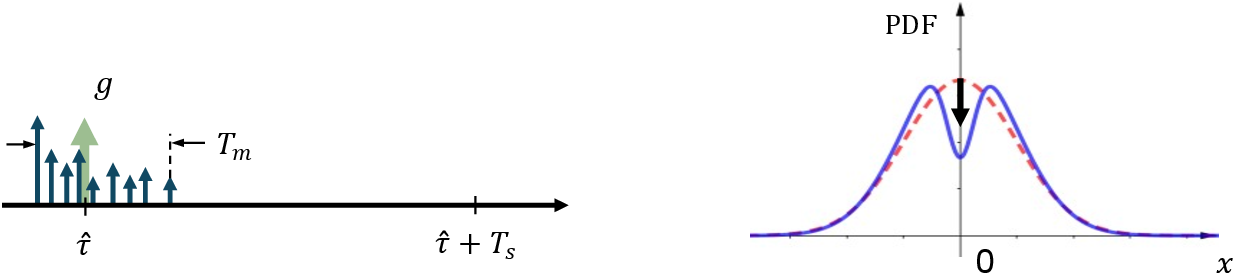}}
	\caption{On the left is a depiction of a typical multipath delay profile of a mediumband channel, where $\hat{\tau}$ is the time offset and also the location of the fading factor denoted by $g$ and shown in light GREEN. Typically, it is the MPCs further away from this fading factor that contribute the most to the effect of deep fading avoidance in $g$. On the right is a PDF of the desired fading factor (only the real part) of mediumband wireless communication in NLoS propagation. The Gaussian PDF is also shown for comparison.}\label{fig:fig01}
	\vspace{0mm}
\end{figure*}

Mediumband wireless communication is neither a modulation scheme nor a coding scheme nor a transmitter precoding scheme nor a receiver combining scheme nor a detection algorithm, nor an anything of that sort. It is an operating condition for wireless communication systems. More precisely, mediumband wireless communication refers to digital wireless communication through mediumband channels, which is a class of channels represented by the GREEN region on the $T_mT_s$-plane in Fig. \ref{fig:fig0}. A wireless communication system, of which $T_m$ and $T_s$ approximately satisfy the condition: $T_m < T_s < 10T_m$, regardless of their carrier frequency, is said to be operating in the mediumband\footnote{Note that this is one of instances, where mediumband channels could occur, but such mediumband channels could also occur in broadband communication as well as discussed in Sec. \ref{sec:mediumband_6G}}.\\
\indent As elaborated in \cite{BasMag2024}, a mediumband channel means a communication scenario where the delay spread is significantly larger in comparison to the symbol period. Consequently, wireless systems operating in this class, or region, experience increased level of inter-symbol-interference (ISI). As depicted on the left in Fig. \ref{fig:fig01}, the multipath delay spread in mediumband is such that the effect of MPCs cannot be fully modelled into a single fading factor, and some leakage receive power left behind by the fading factor gives rise to an additive ISI term. Consequently, the RHS of the baseband equivalent channel in \eqref{eq11} can be tightly modelled for the mediumband region as \cite{Bas2023}:
\begin{align}\label{eq:1}
		r(t) &\approx \overbrace{gs(t-\hat{\tau}) + i(t)}^{r'(t)} + w(t),
\end{align} 
where $g$ is the desired fading factor located at $t=\hat{\tau}$ and $i(t)$ is an additive interference signal emerging from the leakage receive power left behind by the desired signal\footnote{This $\hat{\tau}$ is the time-offset that the RX should be synchronized to. Typically, mediumband wireless communication requires very accurate RX synchronization, specially symbol timing, for optimal performance. If the
RX synchronization in terms of carrier phase and symbol
	timing are suboptimal, the performance should reduce
	gracefully.}. This reduces the quality of the mediumband channel measured in terms of average signal-to-interference-ratio (SIR), which is typically unfavourable. However, the performance of wireless communication in the mediumband is dictated not just by the average SIR, but also the net effect of SIR and another favourable effect known as the “\textit{\textbf{effect of deep fading avoidance}}” in $g$, which can successfully counter the adverse effect of the increased level of ISI. What is this effect of deep fading avoidance?\\
\indent The effect of deep fading avoidance in mediumband channels is such that the probability density function (PDF) of the desired fading factor that arises in mediumband wireless systems (i.e. $g$ in \eqref{eq:1}) exhibits a dip (or hole) at zero signifying the fact that the probability of the desired fading factor being in deep fade is low \cite[Sec. V]{BasMag2024}. Typically, in narrowband communication in non-line-of-sight (NLoS) propagation, the desired fading factor is Gaussian variate, and as a result, there is a peak at zero in the PDF of the desired fading factor, the RED dotted line on the right in Fig. \ref{fig:fig01}. Even in such a highly unfavourable NLoS propagation, if the system is accurately biased to operate in the mediumband, the peak can be converted to a dip at zero giving rise to a bi-modal PDF as shown in BLUE solid line on the right in Fig. \ref{fig:fig01} \cite{BasMag2024}. This effectively means that wireless communication systems can reduce deep fading without doing anything particularly, but by simply operating in the mediumband.\\
\indent Furthermore, the study of mediumband offers us a theoretical foundation to harness the potential of the wireless propagation environment systematically for wider wireless communication. But first, we discuss how to create mediumband channels in practice.   
\section{How To Create Mediumband Channels?}\label{sec:mediumband_howto}
One of the important questions is how to create mediumband channels, or bias a wireless communication system to operate in the mediumband in practice. The starting point to this question is the mediumband constraint that is $T_m < T_s < 10T_m$, which in fact includes two inequalities. Out of these two inequalities, $T_s < 10T_m$ is key\footnote{It is just a historical reason that is behind the other constraint, which is $T_m < T_s$. Because, the BLUE region in Fig. \ref{fig:fig0}, where $T_m \geq T_s$ is traditionally considered as the broadband region. Note also that, as we discuss in Sec. \ref{sec:mediumband_6G}, basic properties of mediumband will not suddenly cease to exist as $T_m$ increases beyond $T_s$.}. If $T_m$ is extremely small in comparison to $T_s$, the channel falls into the narrowband region, and if $T_m$ is moderately large, the channel falls into the mediumband region. There are two ways to design a wireless system to operate in the mediumband: one is at the design-time and other is at real-time: 
\begin{figure*}[t]
	\centerline{\includegraphics*[scale=0.85]{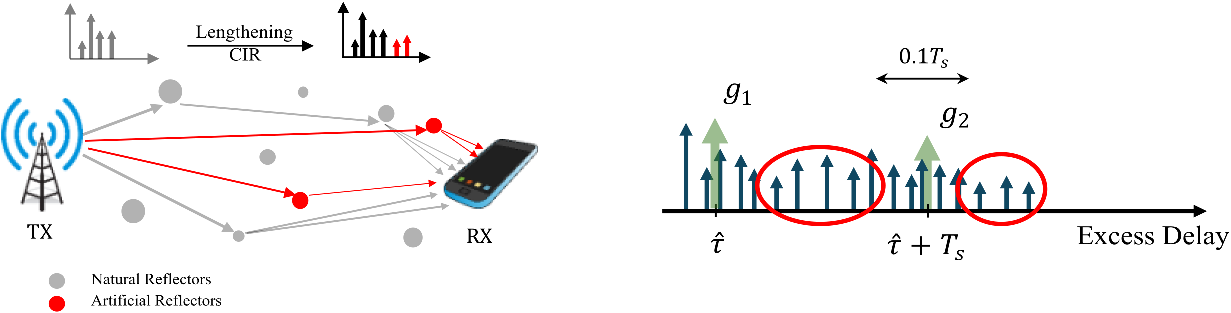}}
	\caption{On the left is a depiction of a wireless link where suitably selected reflecting surfaces artificially lengthen the CIR. Unlike in intelligent reflecting surfaces, herein, no adjustments to the phase of the MPCs and the estimation of the fading factors between nodes and the reflecting surfaces are necessary. On the right is a depiction of a typical multipath delay profile of, as it is called a generalized mediumband channel, where $\hat{\tau}$ is the time offset. The two fading factors, denoted by $g_1$ and $g_2$ shown in light GREEN are modelled to be located at $t=\hat{\tau}$ and $t=\hat{\tau}+T_s$ respectively. Typically, it is the MPCs located further away from these fading factors (circled in RED) that contribute the most to the effect of deep fading avoidance in $g_1$ and $g_2$.}\label{fig:fig02}
	\vspace{0mm}
\end{figure*}
\subsubsection{\textbf{Design-Time Method}}     
Lets assume one wants to design a wireless communication to operate in a particular environment. They should firstly make an estimate about the multipath delay spread, which is $T_m$, of the propagation environment \cite{Sarkar03}. Lets assume, the estimated delay spread is the maximum excess delay, so $\lambda=10$ is reasonable \cite{Bas2023}. With a definitive value for $T_m$ at hand, a symbol period can be chosen to satisfy the mediumband constraint with a desired degree of mediumbandness of their own choosing. Typically, the degree of mediumbandness is measured by the percentage delay spread (PDS) \footnote{The accuracy here could be improved by employing AI models to estimate $T_m$, to choose $T_s$, and also to measure the resulting degree of mediumbandness.}, which is $\text{PDS}=\left(T_m/T_s\right)\times 100\%$. 
%
%
For instance, if $T_m \approx 1\mu s$, designing a wireless communication system with $T_s=2.5\mu s$ ensures that the resulting communication system operates in the mediumband with PDS $=40\%$, and enjoys the deep fading avoidance commensurately. If one wants to increase the degree of mediumbandness further to PDS $=60\%$, the symbol period should be set to $T_s=1.67\mu s$.\\
\indent The PDS is a simple and yet important metric associated with mediumband channels. As PDS increases, the degree of mediumbandness of mediumband channels increases and vice versa. This degree of mediumbandness is such that as it increases many fundamental properties of mediumband channels, such as the depth and width of the dip in $g$, increase favourably \cite{Bas2023,BasMag2024}. 
\subsubsection{\textbf{Real-Time Method}} \label{sec:RT}    
This method is applicable to already designed wireless systems, where $T_s$ and $T_m$ are known to not satisfy the mediumband constraint. For simplicity, we herein consider a narrowband communication system, where by definition, $T_m \ll T_s$ (see the RED region in Fig. \ref{fig:fig0}). Exploiting technologies like reflecting surfaces, as shown on the left in Fig. \ref{fig:fig02}, one can introduce new MPCs to the multipath delay profile lengthening the delay profile, which in turn increases $T_m$\footnote{Note that new MPCs not only increases delay spread, but also contribute favourably to the strength of the desired fading factor too.}. If this increase in $T_m$ is sufficiently large and satisfies the mediumband constraint, the system can be regarded as operating in the mediumband, and enjoy the benefits like deep fading avoidance that is commensurate with the degree of mediumbandness of the resulting mediumband channel\footnote{According to the theory of mediumband, appropriately adjusting $T_m$ in comparison to $T_s$ is all that is needed to change the statistics favourably, where any fine adjustments to the MPCs, such as the phase rotations at reflecting surfaces, are not necessary.}. The system's degree of mediumbandness is dependent on the value of the resulting $T_m$ in comparison to $T_s$. Note that, reflecting surfaces improve the quality of communication not by increasing channel quality metrics like SNR, but by changing the statistics of the desired fading factor. It is important to understand that, in this scenario, the system operates in the mediumband only as long as the reflecting surfaces create new MPCs. Immediately these MPCs cease to exist, $T_m$ falls back to its original value, and in turn, the system falls back to a narrowband system.\\
\indent It is also important to note here that, unlike in intelligent reflecting surfaces \cite{Swindlehurst22}, herein, no adjustments to the phase of the MPCs and the estimation of the channels
to and from the reflecting surfaces may not be necessary. It is anticipated that such operational simplifications for reflecting surfaces, when operating alongside mediumband wireless communication, would give rise to a more compelling and robust use cases, than the highly limited use cases currently being widely studied, for reflecting-surface-enabled wireless communication too. How can we extrapolate these to 6G?  
\vspace{-1mm}
\section{Mediumband Channels In The Context of 6G}\label{sec:mediumband_6G}
Future 6G would include both narrowband and broadband communication. The users (or devices) that has stringent requirements in terms of energy efficiency and reliability, but require low data rates, would likely use narrowband communication, while, users that have more emphasis on data rate will use broadband communication. Both type of communications could benefit from operating in the mediumband. 
\subsubsection{\textbf{Narrowband}}
Narrowband communication means a setup with $T_s \gg T_m$ or at least $T_s > 10T_m$ (see the RED region in Fig. \ref{fig:fig0}). If this communication is designed or biased to operate in the mediumband, that means ensuring $T_s < 10T_m$. For a given $T_m$, this effectively means a communication system with lower symbol period or higher signalling rate which in turn gives rise to higher data rate. Also, due to the effect of deep fading avoidance, new mediumband channel would be more reliable too \cite[Figs. 7 and 8]{BasMag2024}. Therefore, if every narrowband communication link is converted in to a mediumband channel, either at the design stage or in real-time, that would be a ``win-win'' situation.    
\subsubsection{\textbf{Broadband}}\label{mediumband_broadband}
Typically broadband channels are defined in terms of multiple fading factors or coefficients or taps, and by employing IFFT and FFT blocks at the TX and the RX respectively, could be converted to ``\textit{a collection of parallel narrowband channels}'' as in orthogonal frequency-division multiplexing (OFDM). In light of mediumband, we can envision a broadband communication scenario for 6G, where 6G broadband waveforms would, rather than a collection of narrowband waveforms, be a collection of parallel mediumband waveforms, where each sub-channel is capable of handling more interference, and the frequency domain fading factors in every sub-channel also exhibit deep fading avoidance. How does that work?\\
\indent Consider the multipath delay profile shown on the right in Fig. \ref{fig:fig02}, where $T_m \approx 1.2T_s$, which falls in the BLUE region on the $T_mT_s$-plane. This profile is typically modelled in terms of two taps (say $g_1$ and $g_2$) separated by $T_s$, where $g_1$ and $g_2$ are known as the time-domain channel coefficients. As analytically shown in \cite{Bas2023}, if we can introduce MPCs in between $g_1$ and $g_2$, the theory of mediumband ensures that the deep fading avoidance could be induced in $g_1$ and $g_2$. In light of the fact that frequency domain fading coefficients in OFDM are nothing but a weighted sum of the time domain fading factors, frequency domain fading coefficients also exhibit deep fading avoidance. Importantly, it is the MPCs located further away from the location of fading factors (circled in RED) that contribute the most to the effect of deep fading avoidance in $g_1$ and $g_2$. Also, it is the location and the strength of these MPCs with respect to the location and strength of $g_1$ and $g_2$, but not their phase, that is more important for the deep fading avoidance (see the right in Fig. \ref{fig:fig02} for a depiction).              
\section{Further Remarks On Inducing MPCs}
\begin{figure*}[t]
	\centerline{\includegraphics[scale=0.95]{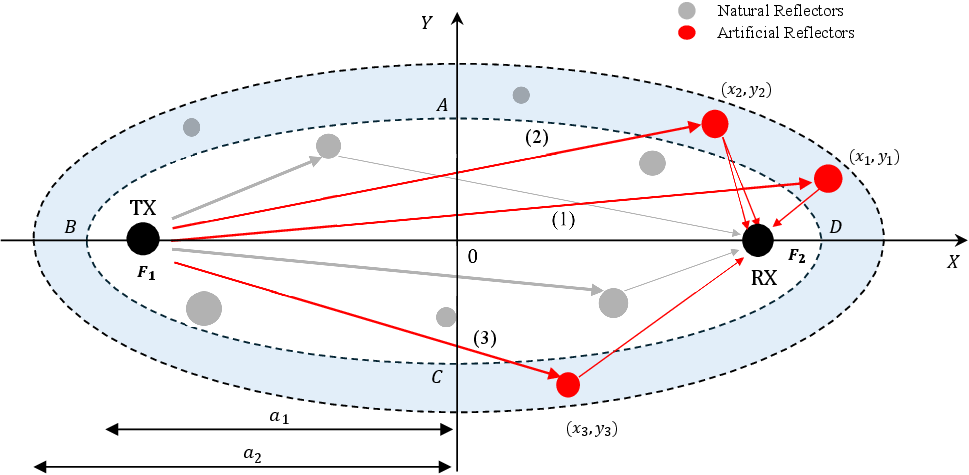}}
	\caption{Top view of a mediumband wireless communication system aided by reflecting surfaces, where artificial reflectors located between two ellipses introduce three MPCs to lengthen the multipath delay profile. Simple geometric arguments ensure that their corresponding propagation delays satisfy: $\frac{2a_1}{c} \leq \tau_1, \tau_2, \tau_3 \leq \frac{2a_2}{c}$, where $c$ is the speed of light, and $2a_1$ and $2a_2$ are the lengths of the semi-major axes of the inner and outer ellipses. Importantly, the precise knowledge of the location of reflectors: $(x_i,y_i)$ are not needed here. For graphical simplicity, it is ellipses that we considered here, but in practice, considering ellipsoids with corresponding geometric arguments would be more appropriate.}\label{fig:fig24}
	\vspace{0mm}
\end{figure*}
Despite the growing body of research, the intelligent reflecting surfaces face major challenges in terms of CSI acquisition, broadband operation, and deployment strategies while being compliant with interoperability and regulatory requirements \cite{Mayur24}. In light of such significant technical and regulatory challenges, mediumband seems to offer reflecting surfaces a viable and robust way for deployment in near future like in 6G as a technology that assist mediumband wireless communication. When operating as an assisting technology for mediumband, reflecting surfaces can bypass stringent CSI requirements and enable broadband communication as outlined in Sec. \ref{mediumband_broadband}. As shown in Sec. \ref{sec:RT}, reflecting surfaces can introduce MPCs into the CIR in order to bias a given communication link to be operated in the mediumband. Furthermore, we may be able to devise simple geometric arguments to choose appropriate reflecting surfaces with manageable signalling overhead.\\
\indent Fig. \ref{fig:fig24} shows the top view of a typical propagation environment, where TX and RX locate on left and right focal points of the ABCD ellipse respectively. The geometry of ellipse ensures that the corresponding propagation delays of the strongest MPCs arising from single reflections occurred within this ellipse would be less than $2a_1/c$, where $c$ is the speed of light and $2a_1$ is the length of the semi-major axis of ABCD. Furthermore, the corresponding propagation delays of the reflectors in RED located between two ellipses satisfy: $2a_1/c \leq \tau_1, \tau_2, \tau_3 \leq 2a_2/c$, where $2a_2$ is the length of the semi-major axis of the outer ellipse. Importantly, the precise knowledge of the location of reflectors: $(x_i,y_i)$ are not needed here. For graphical simplicity, it is ellipses that we considered in this discussion, but in practice, considering ellipsoids with corresponding geometric arguments would be more appropriate.  
\begin{figure*}[t]
	\centerline{\includegraphics[scale=0.86]{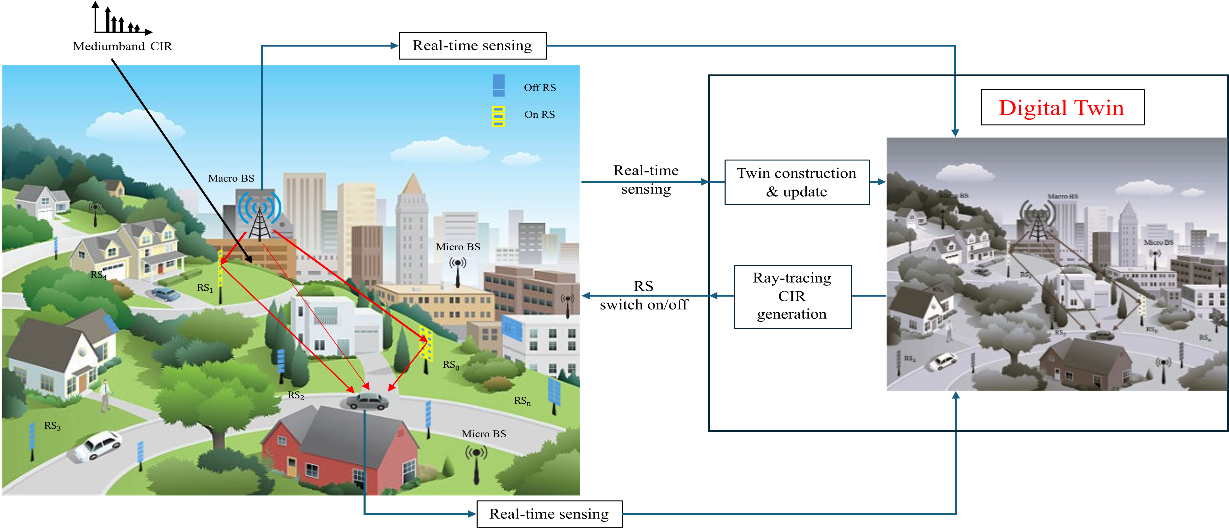}}
	\caption{Depiction of a typical TX-RX pair in the envisioned physical layer architecture for 6G and beyond wireless communication networks, where real-time sensing, ray-tracing, reflecting surfaces (RSs) and AI collaborate smoothly to bias a wireless communication link to operate in the mediumband, and to operate optimally. In the depiction, two reflecting surfaces assist the communication actively to create a mediumband channel, whereas many such RSs scattered in the environment do not engage with the given communication, but may be engaging with other communications.}\label{fig:fig25}
	\vspace{0mm}
\end{figure*}
\section{Mediumband With AI, Sensing AND Digital-Twin}	
The sophistication and robustness of mediumband-based 6G and beyond physical layer can be further improved by the amalgamation of mediumband with sensing, digital twin, and AI. As shown in Fig. \ref{fig:fig25}, the real world and the digital world can be combined, where reflecting surfaces, AI, real-time sensing and ray tracing collaborate smoothly to achieve the objective that is to drive the constituent links in the network to operate in the mediumband and to operate optimally. We may also use the approach known as anticipatory networks, where the behaviour
of users and the status of the network is constantly sensed or predicted, and the predicted information is exploited to make decisions on communication \cite{Hou21}. How might that work in the context of mediumband in the physical layer?\\
\indent On the left in Fig. \ref{fig:fig25} is the real world consisting of a TX and RX pair alongside many other sensing and/or communicating nodes and reflecting surfaces. These nodes, including possibly the given TX-RX pair, continuously sense the environment around them and send the data to the digital twin shown on the right \cite{Alkhateeb23}. In the digital world, using ray-tracing, CIR is estimated with high accuracy and AI may be used to decide if the estimated CIR falls in the region of mediumband or not. If it is, the communication is approved and this message is transmitted to the real world. If the CIR does not satisfy the mediumband constraints, the relevant reflecting surfaces that should be switched on/off in order to alter the CIR appropriately is identified in the digital domain, and in real-time conveyed back to the real world, where the decision will be implemented. If there is a relative movement between TX and RX, the reflecting surfaces that are switched on/off could be updated as appropriately.\\ 
%
%
\textit{\textbf{Role of AI:}} In this paper, we identified areas where AI could assist, or fine-tune algorithms for, mediumband wireless communication and 6G within the boundaries of the mediumband framework. Those include estimating $T_m$, choosing $T_s$ to achieve a certain deep fading avoidance in $g$, estimating the degree of mediumbandness, and symbol timing synchronization at the RX that is estimating $\hat{\tau}$. Furthermore, the mediumband wireless communication at the link-level may further be optimized by using deep learning-based AI techniques for the end-to-end physical layer operation as described in \cite{Qin19}. Importantly, when AI is made to operate within the boundaries of the mediumband framework \cite{Bas2023,BasMag2024}, we may be able to reduce, or perhaps eliminate, the potential technical debts and unknown dependencies that frequently arise in AI algorithms.           
\section{Why Mediumband May Be The Common Denominator In 6G PHY?}	
In addition to the ever present noise, that is $w(t)$ in \eqref{eq:1}, the deep fading is the most prevalent and most consequential effect, or the heaviest burden, in digital wireless communication. The deep fading, an effect caused by the propagation environment, drags the performance of wireless communication down on many fronts. For instance, in conventional narrowband and broadband communication in NLoS propagation, deep fading causes extensive bit errors at RXs decreasing the reliability, and reduces the energy efficiency draining the battery of wireless TXs. Researchers have used various techniques like channel coding and multiple antennas to circumvent this problem \cite{Alkhateeb20, Costello07}.\\
\indent In coding, redundant information is added to the transmit signal to assist the information decoding at the RX, but at the cost of additional energy, lower data rate and added decoding complexity. In multi-antenna techniques, among other things, RX uses multiple antennas to capture more power in order to enhance the strength of the receive signal, which is at the cost of more hardware, higher signal processing burden and additional energy consumption. However, if a wireless link, whether it be a link in sub-6GHz massive MIMO system or in a upper mid-band cell free MIMO network or in a sub-THz wireless personal area network (WPAN) or in a non-terrestrial-network (NTN) or in a cognitive radio system or in a relay communication system or in a narrowband low-power wide area network (LPWAN) or in a wireless ad-hoc network, is successfully designed to operate in the mediumband, it experiences significantly low deep fading with no additional resource requirement. Basically, constituent links in the network being made to operate in the mediumband is analogous to lifting a heavy burden from the network, which is expected to have a favourable knock-on effect on innumerable aspects.
\vspace{0mm}
\section{Conclusion}
In this paper, we articulated a vision for a unified 6G and beyond physical layer where emerging technologies—such as reflecting surfaces, digital twins, precise 3D maps, sensing, ray-tracing, and advanced machine learning—work collaboratively to operate wireless links in the mediumband, addressing key challenges in wireless communication like deep fading. We outlined how the combined capabilities of these physical layer technologies could synergistically contribute to a more robust, efficient, and resilient 6G network. As computing, AI, communication, and synchronization capabilities evolve, we anticipate that networked devices will dynamically build and update digital twins, enabling the 6G and beyond physical layer to be ``\textit{liberated}'' from major dampening forces like deep fading and to potentially achieve substantial improvements in terms of reliability, spectral efficiency, energy efficiency, robustness, and resilience.
\vspace{0mm}
\balance

%
%
\end{document}